\documentclass[aip,jmp,amsmath,amssymb,floatfix,reprint,twocolumn]{revtex4-1}
\usepackage{amsmath,mathtools}
\makeatletter
\renewcommand*\env@matrix[1][\arraystretch]{%
  \edef\arraystretch{#1}%
  \hskip -\arraycolsep
  \let\@ifnextchar\new@ifnextchar
  \array{*\c@MaxMatrixCols c}}
\makeatother
\usepackage{accents}

\usepackage{pgfplots}
\usepackage{epstopdf}
\usepackage{bibentry,natbib}
\usepackage{sublabel}
\usepackage{subfigure}
\usepackage{accents,amsfonts}
\usetikzlibrary{patterns}

\usepackage{graphicx}
\usepackage{helvet}
\usepackage{color}
\usepackage{url}
\usepackage{bm}
\usepackage{upgreek}

\newcommand{\fff}{\mathbf{f}}
\newcommand{\gggg}{\mathbf{g}}
\newcommand{\hh}{\mathbf{h}}
\newcommand{\ph}{\bm{\upphi}}
\newcommand{\ps}{\bm{\uppsi}}
\newcommand{\hps}{\hat{\ps}}
\newcommand{\rr}{\mathbf{r}}
\newcommand{\qq}{\mathbf{q}}
\newcommand{\hrr}{\hat{\rr}}
\newcommand{\sss}{\mathbf{s}}
\newcommand{\uu}{\mathbf{u}}
\newcommand{\huu}{\hat{\uu}}
\newcommand{\MM}{\mathbf{M}}
\newcommand{\WW}{\mathbf{W}}
\newcommand{\Win}{\WW_{\text{in}}}
\newcommand{\Wout}{\WW_{\text{out}}}

\pgfplotsset{compat=1.12}

\begin{document}
\title{Attractor Reconstruction by Machine Learning}


\author{Zhixin Lu}
 \email{zhixinlu@seas.upenn.edu.}
 \affiliation{Department of Bioengineering, University of Pennsylvania, Philadelphia, PA, 19104}
\author{Brian R. Hunt}%
\affiliation{Institute for Physical Science and Technology, University of Maryland, College Park, MD 20742
}%

\author{Edward Ott}
\affiliation{Institute for Research in Electronics and Applied Physics, University of Maryland, College Park, MD 20742
}%

%
%
%
%
%
%
%
\begin{abstract}
A machine-learning approach called ``reservoir computing'' has been used successfully for short-term prediction and attractor reconstruction of chaotic dynamical systems from time series data. We present a theoretical framework that describes conditions under which reservoir computing can create an empirical model capable of skillful short-term forecasts and accurate long-term ergodic behavior. We illustrate this theory through numerical experiments. We also argue that the theory applies to certain other machine learning methods for time series prediction. 
\end{abstract}
%
%
%
%
%
%
%
%
%
%
%

\maketitle
%
%
%
%
%
%
%
%
%
\begin{quotation}
A long-standing problem is prediction and analysis of data generated by a chaotic dynamical system whose equations of motion are unknown. Techniques based on delay-coordinate embedding have been successful for sufficiently low-dimensional systems. Recently, machine-learning approaches such as reservoir computing have shown promise in treating a larger class of systems. We develop a theory of how prediction with reservoir computing or related machine-learning methods can ``learn'' a chaotic system well enough to reconstruct the long-term dynamics of its attractor.
\end{quotation}

\section{\label{sec:introduction}Introduction}

Reservoir computing \cite{jaeger2001echo,maass2002real,jaeger2007echo,lukovsevivcius2009reservoir}
is a machine-learning approach that has demonstrated success at a variety of tasks, including time series prediction
\cite{jaeger2004harnessing,parlitz2005,wyffels2010,pathak2017using} and inferring unmeasured variables of a dynamical system from measured variables \cite{lu2017reservoir,zimmermann2018}.  In this approach, a ``reservoir'' is a high-dimensional, non-autonomous (driven) dynamical system, chosen independently of the task.  A particular task provides an input time series, and the reservoir state as a function of time is regarded as a ``raw'' output time series, which is post-processed to fit the task.  The post-processing function is determined, typically by linear regression, from a limited-time ``training'' data set consisting of the desired output time series for a given input time series.

Reservoir computing can be performed entirely in software, typically with an artificial neural network model, or with a physical reservoir; examples of the latter include a bucket of water \cite{fernando2003pattern}, an electronic circuit with a time delay \cite{appeltant2011information}, a field-programmable gate array (FPGA) \cite{haynes2015reservoir}, an optical network of semiconductor lasers \cite{brunner2016all}, and an optic-electronic phase-delay system \cite{larger2017high}.  Other machine-learning techniques, including deep learning \cite{lecun2015deep,goodfellow2016deep}, attempt to optimize internal system parameters to fit the training data; doing so requires a mathematical model of the machine-learning system.  By contrast, reservoir computing does not require a model for the reservoir, nor the ability to alter the reservoir dynamics, because it seeks only to optimize the parameters of the post-processing function.  The ability to use a physical reservoir as a ``black box'' allows for various potential advantages over other machine-learning techniques, including greatly enhanced speed.

In this article, we consider the task of predicting future measurements from a deterministic dynamical system, whose equations of motion are unknown, from limited time series data.  We describe a general framework that includes the reservoir computing prediction method proposed by Jaeger and Haas \cite{jaeger2004harnessing}.  With appropriate modifications, the same framework applies to other machine-learning methods for time series prediction (including an LSTM approach \cite{vlachas2018}), as we discuss further in Sec.~\ref{sec:discussion}.  We assume the vector $\uu(t)$ of measurements to be a function $\hh$ of the finite-dimensional system state $\sss(t)$,
\begin{equation}\label{eqmeas}
\uu(t) = \hh(\sss(t)).
\end{equation}
For simplicity, we assume that there is no measurement noise, though our discussion below could be modified for the case that Eq.~(\ref{eqmeas}) is an approximation.
We do not assume that $\hh$ is invertible, nor that $\hh$ or $\sss$ is known in practice.  Training data consists of a finite time series $\{\uu(t)\}$ of measurements.  We predict future values of $\uu(t)$ by a sequence of three steps, which we call listening, training, and predicting.

Listening consists of using the training time series as input to the reservoir, which we model as a discrete time deterministic process:
\begin{equation}\label{eqnon}
\rr(t+\tau) = \fff[\rr(t),\uu(t)].
\end{equation}
Here $\rr(t)$ is the reservoir state, $\tau$ is a time increment, and we assume $\fff$ to be a differentiable function.  We emphasize that in practice, a formula for $\fff$ need not be known; only its outputs are used for training and prediction.  For convenience, we assume that the full reservoir state $\rr(t)$ can be measured or computed, though our arguments can be modified easily for the more general case that the reservoir output is a function of its internal state.  We call Eq.~(\ref{eqnon}) the ``listening reservoir''.

Training consists of determining a post-processing function $\hps$ that, when applied to the reservoir output $\rr(t+\tau)$, estimates the next input $\uu(t+\tau)$.  (We view $\hps$ as an approximation to an ``ideal'' post-processing function $\ps$, to be introduced in Sec.~\ref{sec:train}.)  Thus, the goal of training is to find $\hps$ such that $\hps(\rr(t+\tau)) \approx \uu(t+\tau)$, or equivalently,
\begin{equation}\label{eqfit}
\hps(\rr(t)) \approx \uu(t),
\end{equation}
for $t$ large enough that the listening reservoir (\ref{eqnon}) has evolved beyond transient dynamics.  We compute $\hps$ by a fitting procedure, such as linear regression, on the training time series $\{\uu(t)\}$ and the corresponding time series $\{\rr(t)\}$ determined from the listening reservoir (\ref{eqnon}).

Predicting then proceeds by modifying the reservoir to run autonomously with a feedback loop, replacing its input [$\uu(t)$ in Eq.~(\ref{eqnon})] with its post-processed output from the previous time increment:
\begin{equation}\label{eqaut}
\hrr(t+\tau) = \fff[\hrr(t),\hps(\hrr(t))].
\end{equation}
We call Eq.~(\ref{eqaut}) the ``predicting reservoir''.  When initialized (from the listening reservoir state) with $\hrr(t_0) = \rr(t_0)$, iterating the predicting reservoir yields a time series $\{\hps(\hrr(t_0+\tau)), \hps(\hrr(t_0+2\tau)), \ldots\}$ of predictions for future measurements $\{\uu(t_0+\tau), \uu(t_0+2\tau), \ldots\}$.  (Our notation reflects the fact that for $t > t_0$, the predicting reservoir state $\hrr(t)$ estimates the state $\rr(t)$ that would result from evolving the listening reservoir (\ref{eqnon}) with the future measurements.)

The reservoir prediction method we have described has been shown to produce successful short-term forecasts for a variety of dynamical systems \cite{jaeger2004harnessing,wyffels2010,pathak2017using}.  If the system has a chaotic attractor, then, as for any imperfect model, the prediction error $\|\hps(\hrr(t)) - \uu(t)\|$ cannot remain small for $t \gg t_0$.  However, in some cases, the long-term time series $\{\hps(\hrr(t))\}$ continues to behave like the measurements from a typical trajectory on the attractor, and in this sense the predicting reservoir (\ref{eqaut}) approximately reproduces the ergodic properties of the dynamical system that generated the measurements \cite{pathak2017using}.  We refer to this ability, often called attractor reconstruction, as replication of the ``climate''.

In this article, we develop and illustrate a theory of how reservoir prediction is able to ``learn'' the dynamics of a system well enough to produce both accurate short-term forecasts and accurate long-term climate.  We make use of the notion of \emph{generalized synchronization} \cite{afraimovich1986stochastic,pecora1990synchronization,rulkov1995generalized,kocarev1996}, which in our context means that the reservoir state $\rr(t)$ becomes asymptotically a continuous function $\ph$ of $\sss(t)$,
in the limit that the listening reservoir (\ref{eqnon}) is run infinitely long.  In Sec.~\ref{sec:Hunt}, we argue that the following four conditions are sufficient for both short-term prediction and attractor/climate replication.
\begin{enumerate}

\item
The listening reservoir (\ref{eqnon}) achieves generalized synchronization with the process $\{\sss(t)\}$, so that $\rr(t) \approx \ph(\sss(t))$ for a continuous function $\ph$, within the time interval covered by the training time series.

\item
The synchronization function $\ph$ is one-to-one, or at least carries enough information about its input to recover $\uu(t) = \hh(\sss(t))$ from $\ph(\sss(t))$.

\item
Training is successful in finding a function $\hps$ such that Eq.~(\ref{eqfit}) holds, or equivalently in view of generalized synchronization, that $\hps(\ph(\sss(t))) \approx \hh(\sss(t))$.

\item
The attractor approached by the listening reservoir is also stable for the predicting reservoir (\ref{eqaut}).

\end{enumerate}
Conditions 1--3 enable short-term prediction. Condition~4 ensures that the climate established by generalized synchronization of the listening reservoir is preserved when its input is replaced by a feedback term to form the predicting reservoir.  One of the main points of Sec.~\ref{sec:Hunt} is to precisely formulate the stability condition described in Condition~4.

We remark that generalized synchronization of the listening reservoir \cite{parlitz2005,zimmermann2018} is related to the
``echo state property'' \cite{jaeger2001echo,yildiz2012}, which states that an infinite history of inputs $\{\uu(t-\tau), \uu(t-2\tau), \ldots\}$ uniquely determines $\rr(t)$, subject to the condition that the trajectory $\{\rr(t)\}$ is bounded.  Indeed, if $\{\sss(t)\}$ is a trajectory of an invertible dynamical system, then the past inputs are functions of $\sss(t)$, so the echo state property implies that if the listening reservoir (\ref{eqnon}) has run for an infinite period of time in a bounded domain, then $\rr(t)$ is a function of $\sss(t)$ [though it does not imply that this function is continuous].  We believe that for the reservoir prediction method we described, it is desirable (though not strictly necessary) to have the echo state property and generalized synchronization.  In Sec.~\ref{sec:synch}, we show why both properties hold if the listening reservoir is uniformly contracting as a function of $\rr$, and that we can quantify the amount of transient time it takes for the reservoir to achieve the approximation $\rr(t) \approx \ph(\sss(t))$ of Condition~1.

Conditions~2 and 3 are significantly more difficult to ensure \emph{a priori}.  In Sec.~\ref{sec:train}, we argue why it is plausible that these conditions can be achieved.  In Secs.~\ref{sec:attractor} and \ref{sec:lyapunov}, we describe the consequences of Conditions~1-3 for short-term prediction, and formulate more precisely the stability criterion of Condition~4 that determines whether the correct attractor and climate are approximately reproduced by the long-term dynamics of the predicting reservoir (\ref{eqaut}).  In Sec.~\ref{sec:compute}, we describe how a model for the reservoir dynamics can be used to compute Lyapunov exponents that reflect climate stability.

In Sec.~\ref{sec:RCN}, we give examples of short-term state and long-term climate predictions using the Lorenz equations as our input system.  In addition to a case where the climate is approximated well, we show a case where the predicted climate is inaccurate, though the short-term forecast is still reasonably accurate.  We compute the Lyapunov exponents of the predicting reservoir (\ref{eqaut}), and show that the transition from accurate climate to inaccurate climate corresponds to a Lyapunov exponent crossing zero.  When this Lyapunov exponent is positive but close to zero, the reservoir prediction remains close to the correct climate for a transient period, and we relate the average duration of this transient to the value of the Lyapunov exponent.

\section{\label{sec:Hunt}Theory}

\begin{figure}[htbp]
	\centering
	\includegraphics[width=0.4\textwidth]{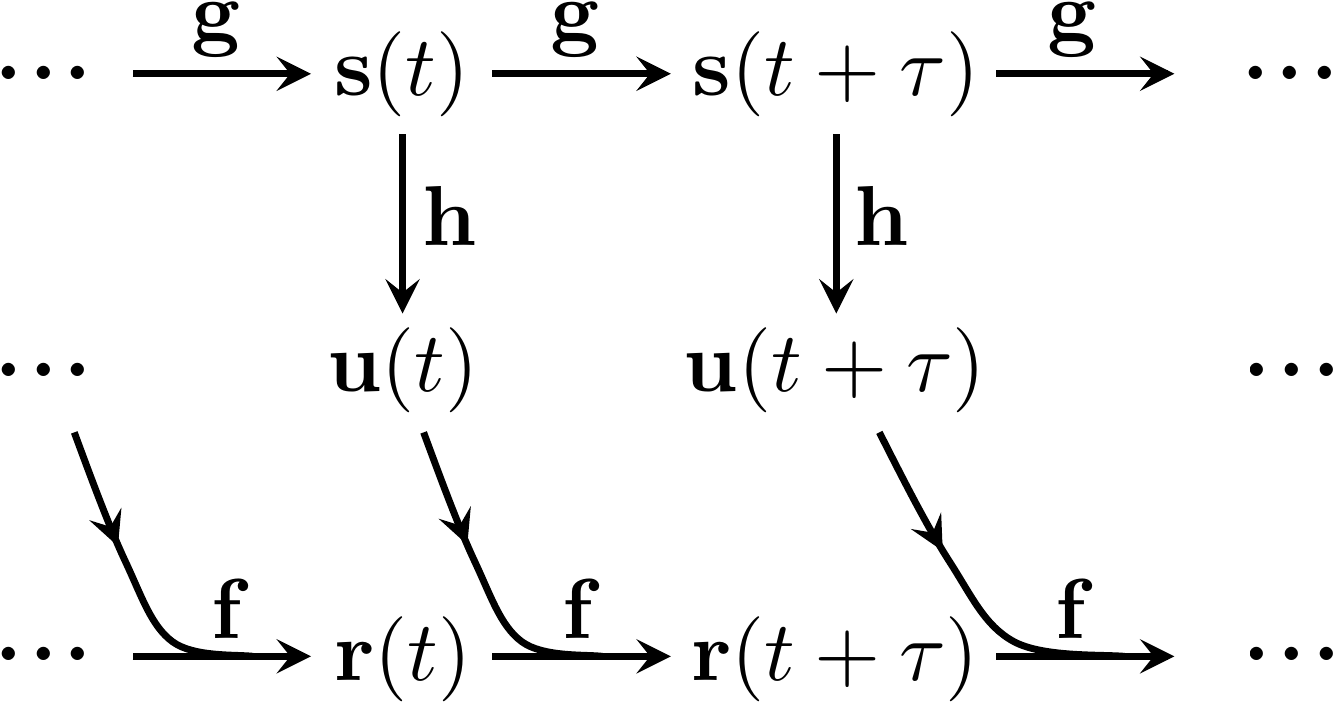}
	\caption{Drive-response system dynamics, with the drive state $\sss(t)$ coupled to the listening reservoir state $\rr(t)$ through the measurement vector $\uu(t)$.}
\label{fig:new_figure_1_of_sec_2}
\end{figure}

We consider the application of the reservoir prediction method described in the introduction to a time series $\{\uu(t)\}$ that is a function $\hh$ of a trajectory $\{\sss(t)\}$ of the dynamical system
\begin{equation}\label{eqinp}
\sss(t+\tau) = \gggg(\sss(t)),
\end{equation}
where $\gggg$ is differentiable and invertible, and we assume that $\sss(t)$ evolves on a bounded attractor $A$.  In preparation for training and prior to prediction, the reservoir state $\rr(t)$ evolves according to the listening reservoir (\ref{eqnon}).  The system described by Eqs.~(\ref{eqinp}) and (\ref{eqnon}), coupled by Eq.~(\ref{eqmeas}), is often called a drive-response, skew-product, or one-way coupled system.  The coupled system dynamics are illustrated by Fig.~\ref{fig:new_figure_1_of_sec_2}.  We next consider the evolution of the coupled system as $t \to \infty$.

\subsection{Listening and Generalized Synchronization}\label{sec:synch}

The goal of training can be regarded as finding a post-processing function $\hps$ such that $\hps(\rr(t))$ is in approximate identical synchronization \cite{pecora1990synchronization} with $\uu(t) = \hh(\sss(t))$,
when $\rr(t)$ is evolved with the listening reservoir (\ref{eqnon}).  The desired relationship $\uu(t) \approx \hps(\rr(t))$ can also be thought of as approximate generalized synchronization between $\uu(t)$ [or the underlying state $\sss(t)$] and $\rr(t)$.  The existence of such a relationship would be implied by stochastic synchronization \cite{afraimovich1986stochastic}, which in our context means a one-to-one correspondence between $\rr(t)$ and $\sss(t)$ in the limit $t\to\infty$.  However, in drive-response systems, the definition of \emph{generalized synchronization} \cite{rulkov1995generalized,kocarev1996} requires only that the
response state be asymptotically a function of the drive state: in our case, that there is a continuous function $\ph$ such that $\rr(t) - \ph(\sss(t)) \to 0$ as $t \to\infty$.  The existence of such a $\ph$ is typically easier to establish than its invertibility.  Next, we describe conditions on the reservoir system $\fff$ that guarantee generalized synchronization.

Though weaker conditions are possible, we assume uniform contraction for $\fff$, as is often the case in practice.  By \emph{uniform contraction}, we mean that there is some $\rho < 1$ such that for all $\rr_1$, $\rr_2$, and $\uu$ we have that $|\fff(\rr_1,\uu)-\fff(\rr_2,\uu)| < \rho |\rr_1-\rr_2|$.  It then follows that two trajectories $\{\rr_1(t),\uu(t)\}$ and $\{\rr_2(t),\uu(t)\}$ of (\ref{eqnon}) with the same input time series approach each other exponentially: $|\rr_1(t) - \rr_2(t)| \leq |\rr_1(0) - \rr_2(0)| \rho^{t/\tau}$.  Thus, for a given input time series $\{\uu(t)\}$, the reservoir state $\rr(t)$ is asymptotically independent of its initial state; this is essentially what Jaeger \cite{jaeger2001echo} called the ``echo state property''.  Furthermore, because $\gggg$ is invertible and $A$ is bounded, and due to results of Hirsch, Pugh, and Shub \cite{hirsch1970,hirsch1977} (a direct proof is given by Stark \cite{stark1997}), uniform contraction implies generalized synchronization, as defined above.  (In general, the synchronization function $\ph$ cannot be determined analytically from $\fff$, $\gggg$, and $\hh$.)  A weaker form of generalized synchronization can also be guaranteed \cite{stark1997} from the non-uniform contraction implied by negative conditional Lyapunov exponents.

We remark that if the listening reservoir (\ref{eqnon}) is uniformly contracting, then $\rr(t) - \ph(\sss(t))$ converges to zero exponentially.  If the designer of the reservoir can guarantee a specific contraction rate $\rho$, this determines the convergence rate, so that the amount of transient time needed to make the approximation $\rr(t) \approx \ph(\sss(t))$ accurate can be known in practice.

Generalized synchronization implies that the set of $(\sss,\rr)$ such that $\sss$ is on its attractor $A$ and $\rr = \ph(\sss)$ is an attractor for the drive-response system given by Eqs.~(\ref{eqinp}), (\ref{eqmeas}), and (\ref{eqnon}).  Below we will use the fact that this set is invariant: $\rr(t) = \ph(\sss(t))$ implies $\rr(t+\tau) = \ph(\sss(t+\tau))$.

\subsection{Training}
\label{sec:train}

Recall that training seeks a function $\hps$ that predicts the current measurement vector $\uu(t)$ from the current listening reservoir state $\rr(t)$ [which is computed from past measurements], and that when generalized synchronization is achieved, accuracy of this prediction is equivalent to $\hps(\ph(\sss(t))) \approx \hh(\sss(t))$.  For the rest of Section~\ref{sec:Hunt}, we assume that there is a function $\ps$ defined on $\ph(A)$ such that $\ps(\ph(\sss)) = \hh(\sss)$ for all $\sss$ in $A$.  This assumption means that in the asymptotic limit of generalized synchronization, the listening reservoir state $\rr(t) = \ph(\sss(t))$ uniquely determines $\uu(t) = \hh(\sss(t))$.  The goal of training can then be described as finding a function $\hps$ defined on the state space of the reservoir that approximates $\ps$ on the set $\ph(A)$.  We summarize our notation in Table~\ref{tab:notation}.

\begin{table}[htbp]
\begin{tabular}{lcl}
\hline
\hline
\multicolumn{3}{c}{Dynamical System to be Predicted}\\
$\sss(t)$ && System state \\
$\gggg : \sss(t) \to \sss(t+\tau)$ && System evolution \\
$A$ && Attractor for $\sss(t)$ \\
\hline
\multicolumn{3}{c}{Measurements}\\
$\uu(t)$ && Measurement vector \\
$\hh : \sss(t) \to \uu(t)$ && Measurement function \\
\hline
\multicolumn{3}{c}{Reservoir}\\
$\rr(t)$ && Listening reservoir state \\
$\fff : [\rr(t),\uu(t)] \to \rr(t+\tau)$ && Listening reservoir evolution \\
$\hrr(t)$ && Predicting reservoir state \\
$\huu(t) = \hps(\hrr(t))$ && Predicted measurements \\
$\fff : [\hrr(t),\huu(t)] \to \hrr(t+\tau)$ && Predicting reservoir evolution \\
\hline
\multicolumn{3}{c}{Generalized Synchronization}\\
$\ph : \sss \to \rr$ for $\sss$ in $A$ && Synchronization function \\
$\ps : \rr \to \uu$ for $\rr$ in $\ph(A)$ && Ideal post-processing function \\
$\hps : \hrr(t) \to \huu(t)$ && Actual post-processing function \\
\hline
\hline
\end{tabular}
\caption{Summary of Notation}\label{tab:notation}
\end{table}

Though the existence of $\ps$ is not strictly necessary for the reservoir to make useful predictions, if no such $\ps$ exists, then it seems unlikely that training can successfully achieve the desired approximation $\ps(\ph(\sss(t))) \approx \hh(\sss(t))$, and thus unlikely that $\uu(t)$ can be approximated as a function of the reservoir state during either listening or predicting.  The existence of $\ps$ is guaranteed if $\ph$ is one-to-one on $A$; then $\ps = \hh \circ \ph^{-1}$.  Furthermore, if $\hh$ is one-to-one on $A$ (in other words, the measurements at a given time determine the system state), then $\ph$ must be one-to-one on $A$ in order for $\ps$ to exist.  Thus, we propose that a goal of reservoir design should be to yield a one-to-one synchronization function $\ph$ for a variety of input systems.  In practice, having a sufficiently high-dimensional reservoir may suffice; embedding results\cite{whitney,embedology} imply that if the dimension of the reservoir state $\rr$ is more than twice the dimension of $A$, functions from $A$ to the reservoir state space are typically one-to-one.  We note that in practice, the dimension of $\rr$ must be much larger than twice the dimension of $A$ in order to provide a suitable basis for approximating $\ps$, in the sense described below.

Careful consideration of conditions under which training is successful in determining an accurate approximation $\hps$ to $\ps$ is beyond the scope of our theory.  However, we argue that success is plausible if the training time series is sufficiently long that the trajectory $\{\sss(t)\}$ well samples its attractor $A$, if the dimension of the reservoir state $\rr(t)$ is sufficiently high, and if the dynamics of the coordinates of $\rr(t)$ are sufficiently heterogeneous.  If, for example, training uses linear regression of $\{\uu(t)\} = \{\hh(\sss(t))\}$ versus $\{\rr(t)\}$, then since $\rr(t) \approx \ph(\sss(t))$, the coordinates of the vector-valued function $\ph(\sss)$ can be thought of ``basis functions'' \cite{parlitz2005}; training seeks a linear combination $\hps$ of these basis functions that approximates $\hh(\sss)$ on $A$.  A suitable basis for training (using a linear or nonlinear combination) is plausible if the listening reservoir yields a sufficiently large variety of responses to its input.

\subsection{Prediction and Attractor Reconstruction}\label{sec:attractor}

\begin{figure}[htbp]
	\centering
	\includegraphics[width=0.4\textwidth]{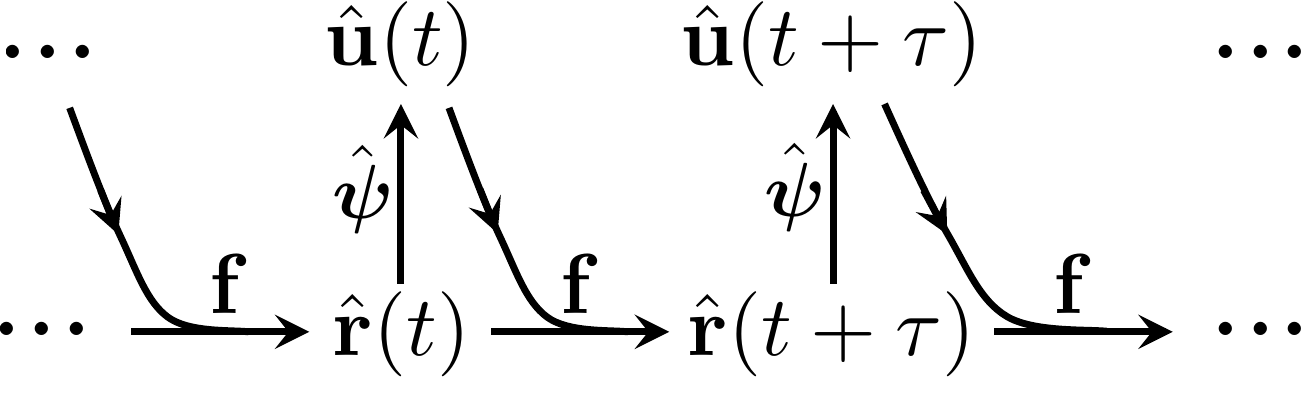}
	\caption{Predicting reservoir dynamics, with the listening reservoir input $\uu(t)$ replaced by the estimate $\huu(t)$ determined from the predicting reservoir state $\hrr(t)$.}
\label{fig:new_figure_2_of_sec_2}
\end{figure}

After training determines the post-processing function $\hps$, prediction proceeds by initializing $\hrr(t_0) = \rr(t_0)$ and evolving $\hrr(t)$ for $t \geq t_0$ according to the predicting reservoir (\ref{eqaut}).  The reservoir state $\rr(t_0)$ is determined by evolving the listening reservoir (\ref{eqnon}) for an interval of time preceding $t_0$; this could be the time interval used for training, or it could be a later time interval that uses inputs $\{\uu(t)\}$ measured after training (we call this feature ``training reusability'' \cite{pathak2018}).  We assume that the listening time preceding $t_0$ is sufficiently long to achieve generalized synchronization, so that $\hrr(t_0) = \rr(t_0) \approx \ph(\sss(t_0))$ is near $\ph(A)$.  For $t \geq t_0$, the predicted value of $\uu(t)$ is
\begin{equation}\label{eqpred}
\huu(t) = \hps(\hrr(t)).
\end{equation}
Figure~\ref{fig:new_figure_2_of_sec_2} depicts the dynamics of the predicting reservoir (\ref{eqaut}).  

Consider now the idealized scenario that our approximations are instead exact relations $\hps = \ps$ on $\ph(A)$, and $\hrr(t_0) = \rr(t_0) = \ph(\sss(t_0))$.  Suppose hypothetically that the measurements $\{\uu(t)\}$ for $t \geq t_0$ (these are the values we want to predict in practice) are available, so that we can evolve both the listening reservoir (\ref{eqnon}) depicted in Fig.~\ref{fig:new_figure_1_of_sec_2}, and the predicting reservoir (\ref{eqaut}) depicted in Fig.~\ref{fig:new_figure_2_of_sec_2}, and compare their outputs.  Then we claim that the two reservoirs agree exactly: $\hrr(t) = \rr(t)$ and $\huu(t) = \uu(t)$ for all $t \geq t_0$.  First notice that $\huu(t_0) = \hps(\hrr(t_0)) = \ps(\ph(\sss(t_0)) = \hh(\sss(t_0)) = \uu(t_0)$.  Then $\hrr(t_0+\tau) = \fff[\hrr(t_0),\huu(t_0)] = \fff[\rr(t_0),\uu(t_0)] = \rr(t_0+\tau)$, and $\rr(t_0+\tau) = \ph(\sss(t_0+\tau))$ due to generalized synchronization.  Similarly, $\huu(t_0+\tau)$ then equals $\uu(t_0+\tau)$, so $\hrr(t_0+2\tau) = \rr(t_0+2\tau) = \ph(\sss(t_0+2\tau))$, etc.  This agreement between the trajectories also shows that $\ph(A)$ is an invariant set for the idealized predicting reservoir
\begin{equation}\label{eqid}
\rr(t+\tau) = \fff[\rr(t),\ps(\rr(t))],
\end{equation}
and that its dynamics, observed through $\ps$, are equivalent to the dynamics of $A$ observed through $\hh$.

Thus, if the time series $\{\uu(t)\}$ of measurements has enough information to reconstruct the attractor $A$, then we can regard $\ph(A)$ and the idealized predicting reservoir (\ref{eqid}) as an exact reconstruction of $A$ and its dynamics.  When the approximation $\hps \approx \ps$ is not exact on $\ph(A)$, the actual predicting reservoir (\ref{eqaut}) is still initialized near $\ph(A)$, but $\ph(A)$ is only approximately invariant.  The better the approximation, the more accurate the predictions $\huu(t) \approx \uu(t)$ will be, at least in the short term.  However, if the system (\ref{eqinp}) that generates the measurements $\{\uu(t)\}$ is chaotic, the prediction error $\|\huu(t) - \uu(t)\|$ will typically grow exponentially as $t$ increases.

Nonetheless, it remains possible that $\huu(t)$ will maintain a climate similar to $\uu(t)$ in the long term.  This will happen if (and practically speaking, only if) the predicting reservoir trajectory $\{\hrr(t)\}$ remains close to $\ph(A)$ for all time, and its attractor has a similar climate to that of the idealized predicting reservoir on $\ph(A)$.  In this sense, climate replication (attractor reconstruction) relies on both state-space stability and structural stability of the predicting reservoir near the idealized reconstructed attractor $\ph(A)$.

Structural stability is difficult to ensure rigorously, but in practice small perturbations of the dynamics near an attractor tend to yield small perturbations to the climate.  Thus, we argue that climate replication is likely if $\ph(A)$, which according to our assumptions is invariant for the idealized predicting reservoir, is also attracting, in the sense described below.

\subsection{Stability and Lyapunov Exponents}\label{sec:lyapunov}

Recall that generalized synchronization implies that the set $\ph(A)$ is attracting for the listening reservoir (\ref{eqnon}), when driven by $\uu(t) = \hh(\sss(t))$ where $\sss(t)$ evolves on $A$.  Whether $\ph(A)$ is attracting for the predicting reservoir is complicated by the fact that it is invariant only in the idealized case $\hps = \ps$, and that $\ps$ is defined only on $\ph(A)$, so that the idealized predicting reservoir (\ref{eqid}) is also defined only on $\ph(A)$.  For its stability to be well-defined, the domain of $\ps$ must be extended to a neighborhood of $\ph(A)$, and whether $\ph(A)$ is attracting depends on how the extension is chosen.

Thus, the suitability of the empirically determined function $\hps$ for climate prediction depends not only on how well it approximates $\ps$ on $\ph(A)$, but also on how it behaves near $\ph(A)$.  For a particular $\hps$, we consider hypothetically a particular extension of $\ps$ such that $\hps \approx \ps$ near $\ph(A)$.  This extension gives the idealized predicting reservoir a full set of Lyapunov exponents on $\ph(A)$, some of which correspond to infinitesimal perturbations tangent to $\ph(A)$ and some of which correspond to infinitesimal perturbations transverse to $\ph(A)$.  Then $\ph(A)$ is attracting if the transverse Lyapunov exponents are all negative, and is unstable if there is a positive transverse Lyapunov exponent.

If the generalized synchronization function $\ph$ is one-to-one and differentiable, then the tangential Lyapunov exponents of the system (\ref{eqinp}) on $A$ are reproduced as the tangential Lyapunov exponents of the idealized predicting reservoir on $\ph(A)$.  Generalized synchronization does not always yield a differentiable $\ph$ \cite{hunt1997,stark1997}, but even when differentiability cannot be guaranteed, it is possible in practice to reproduce much of the Lyapunov spectrum of $A$, including negative Lyapunov exponents in some cases, with a predicting reservoir \cite{pathak2017using}.

We remark that unlike the conditional Lyapunov exponents for a drive-response system (such as the listening reservoir), which correspond to perturbations of the response system state, for the predicting reservoir it is not clear in advance which perturbations correspond to transverse Lyapunov exponents.  However, in a numerical experiment where the equations for the driving system (\ref{eqinp}) and the reservoir are known, the existence or absence of a positive transverse Lyapunov exponent can be inferred by computing all of the positive Lyapunov exponents of the predicting reservoir and eliminating those that are Lyapunov exponents of $A$.

\subsection{Computation of Lyapunov Exponents}\label{sec:compute}

We now describe how to estimate the Lyapunov exponents of the idealized predicting reservoir (\ref{eqid}) on $\ph(A)$, for a particular extension of $\ps$ to a neighborhood of $\ph(A)$, from its empirical approximation $\hps$.  To do so, we assume that we have a formula for $\fff$, so that we can compute its Jacobian matrix.  (We emphasize that we estimate the Lyapunov exponents in order to corroborate the theory we have presented; their computation, and a formula for $\fff$, are not needed for the reservoir prediction method we have described.)  If climate replication is successful, we can simply generate a long trajectory of the predicting reservoir (\ref{eqaut}), and use it to compute the Lyapunov exponents of the trajectory \cite{pathak2017using}.  However, this trajectory cannot be expected to remain close to $\ph(A)$ if the set is unstable.  Nonetheless, if we have a sufficiently long time time series $\{\uu(t)\}$ of measurements, we can estimate the Lyapunov exponents of $\ph(A)$, whether or not it is stable, as follows.

First, we use the time series $\{\uu(t)\}$ to generate a trajectory $\{\rr(t)\}$ of the listening reservoir (\ref{eqnon}); as we have argued, $\rr(t)$ will approach $\ph(A)$ under the conditions for generalized synchronization.  Then along this trajectory, which is an approximate trajectory for the predicting reservoir, we compute Lyapunov exponents using the Jacobian matrix of the predicting reservoir (\ref{eqaut}).

\section{\label{sec:RCN}Numerical Experiments}

In this section, we give examples of short-term state and long-term climate predictions for the Lorenz system \cite{lorenz}, with standard parameter values that yield chaotic trajectories:
\begin{align}
\begin{split}
    dx/dt &= 10(y-x),\\
    dy/dt &= x(28-z) -y,\\
    dz/dt &= xy - 8z/3.
\end{split}\label{eqn:lorenz}
\end{align} 
We consider the case where the measurement function $\hh$ is the identity, so that $\uu(t) = \sss(t) = [x(t),y(t),z(t)]^T$.  For the reservoir, we use an artificial neural network similar to the one used by Jaeger and Haas\cite{jaeger2004harnessing}; our listening reservoir [a continuous-time version of Eq.~(\ref{eqnon})] evolves according to
\begin{equation}
	\frac{d}{dt}\rr(t) = \gamma[-\rr(t)+\tanh(\MM\rr(t)+\sigma\Win\uu(t))],
	\label{eqn:listening_reservoir}
\end{equation}
where $\rr$ is an $N$-dimensional vector, $\gamma$ is a scalar, $\MM$ is an adjacency matrix representing internal network connections.  The matrix $\sigma\Win$ consists of ``input weights''; in our numerical results, we will fix $\Win$ and vary the scalar input strength $\sigma$.  The vector function $\tanh$ is computed by applying the scalar hyperbolic tangent to each coordinate of its input vector.  We compute trajectories of both the Lorenz and reservoir systems using the $4$th order Runge-Kutta method with time step $\tau = 0.001$.  We will show cases where climate replication (attractor reconstruction) succeeds and where it fails, and compare the results with Lyapunov exponents we compute for the predicting reservoir.

\begin{figure}[htbp]
	\centering
	\includegraphics[scale=.5]{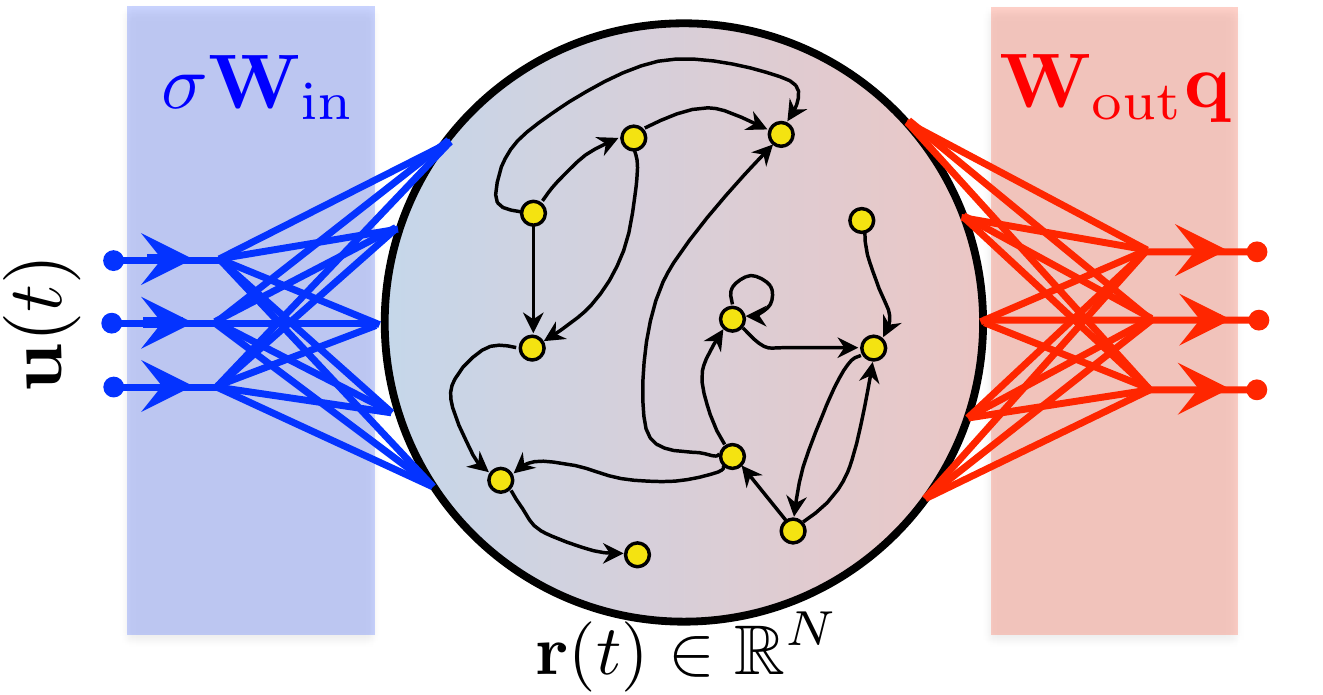}
	\caption{Listening reservoir based on an artificial neural network with $N$ neurons.  The input vector $\uu(t) \in \mathbb{R}^3$ is mapped to the reservoir state space $\mathbb{R}^N$ by the input weight matrix $\sigma\Win$, and the resulting reservoir state is mapped to $\mathbb{R}^3$ by the post-processing function $\hps = \Wout\qq$.}
\label{fig:reservoir1}
\end{figure}

\begin{figure}[htbp]
	\centering
	\includegraphics[scale=.5]{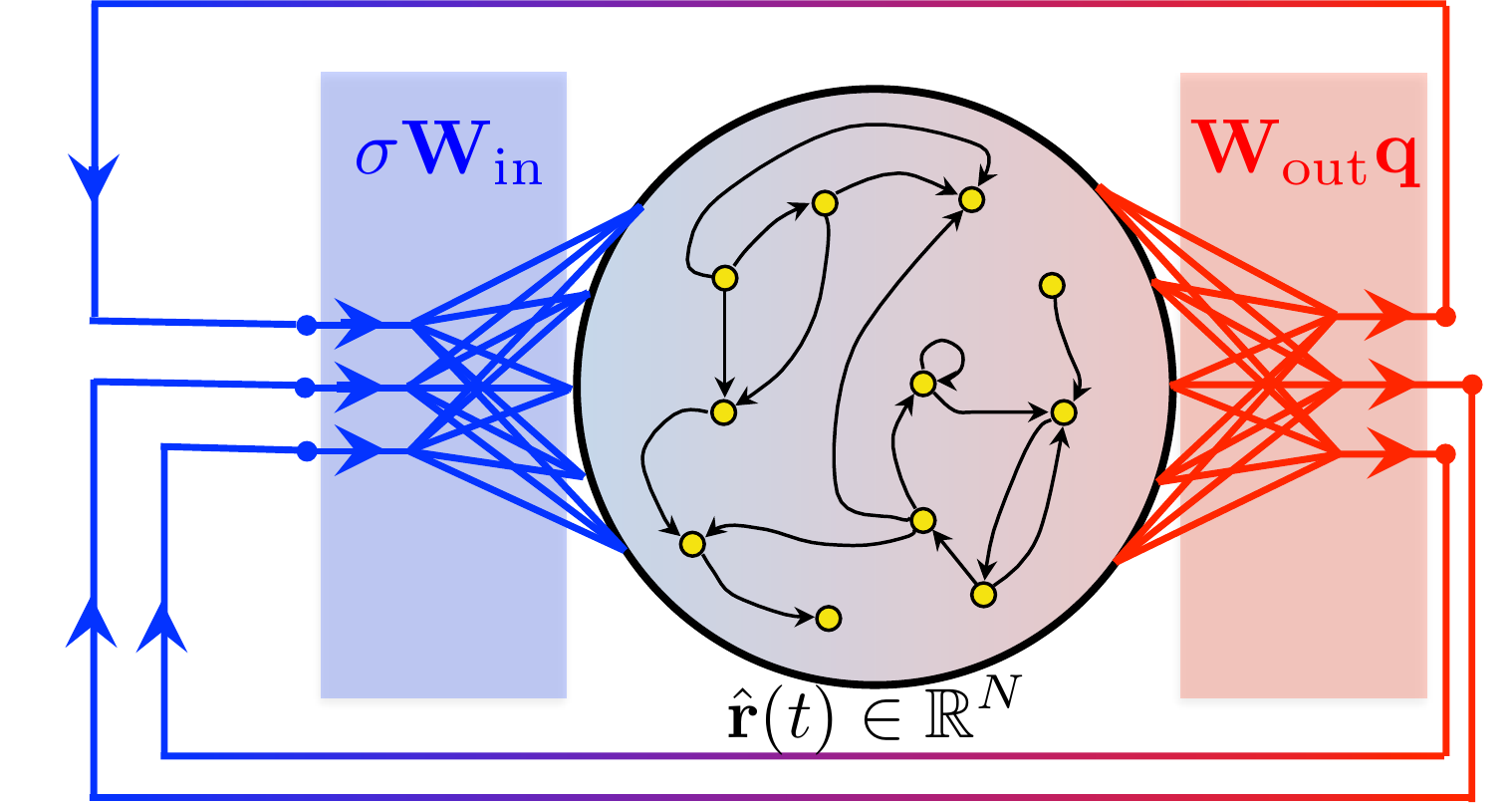}
	\caption{The predicting reservoir replaces the external input of the listening reservoir with the post-processed reservoir output.  The time increment $\tau$ in our discussion represents the amount of time for information to travel once around the feedback loop.}
\label{fig:reservoir2}
\end{figure}

We consider post-processing functions of the form $\hps(\rr) = \Wout\qq(\rr)$, where $\qq(\rr)$ is the $2N$-dimensional vector consisting of the $N$ coordinates of $\rr$ followed by their squares, and the ``output weight'' matrix $\Wout$ is determined by a linear regression procedure described below.  The listening reservoir (\ref{eqn:listening_reservoir}) and the post-processing function are illustrated as an input-output system in Fig.~\ref{fig:reservoir1}.  The goal of training is that the post-processed output $\Wout\qq(\rr(t+\tau))$ based on input up to time $t$ estimates the subsequent input $\uu(t+\tau)$.  Once $\Wout$ is determined, the external input can be replaced in a feedback loop by the post-processed output to form the predicting reservoir, as depicted in Fig.~\ref{fig:reservoir2}.  The predicting reservoir evolves according to
\begin{equation}
	\frac{d}{dt}\hrr(t) = \gamma[-\hrr(t)+\tanh(\MM\hrr(t)+\sigma\Win\Wout\qq(\hrr(t))],
	\label{eqn:predicting_reservoir}
\end{equation}
and the predicted value of $\uu(t)$ is $\huu(t) = \hps(\hrr(t)) = \Wout\qq(\hrr(t))$.

Details of our reservoir implementation are as follows.  The reservoir dimension is $N = 2000$, and we use $\gamma = 10$.  The $N$-by-$N$ adjacency matrix $\MM$ is chosen randomly with sparse Erd\"{o}s-Renyi connectivity and spectral radius $0.9$; specifically, each element is chosen independently to be nonzero with probability $0.02$, nonzero elements are chosen uniformly between $-1$ and $1$, and the resulting matrix is rescaled so that the magnitude of its largest eigenvalue is $0.9$.  The $N$-by-$3$ matrix $\Win$ is chosen randomly so that each row has one non-zero element, chosen uniformly between $-1$ and $1$.  We evolve the Lorenz system and the listening reservoir (\ref{eqn:listening_reservoir}) from time $t = -100$ to $t = 60$, and we discard $100$ time units of transient evolution, so that training is based on $\uu(t)$ and $\rr(t)$ for $0 \leq t \leq 60$.  For training, we constrain the $3$-by-$2N$ matrix $\Wout$ to have only $3N$ nonzero elements, namely the first $N$ elements of its first two rows, and the first $N/2$ and last $N/2$ elements of its third row.  (Thus, we fit the $x$ and $y$ coordinates of the Lorenz state with linear functions of $\rr$, and the $z$ coordinate with a linear combination of the first $N/2$ coordinates of $\rr$ and the squares of the second $N/2$ coordinates; for the Lorenz system, this is advantageous over using a purely linear function of $\rr$ \cite{pathak2017using}.)  Subject to this constraint, we select $\Wout$ so as to minimize the error function
\begin{equation}
\sum_{k=1}^{3000}\Vert\Wout\qq(\rr(0.02k)) - \uu(0.02k)\Vert^2 + \beta \Vert\Wout\Vert^2;
\label{eqn:error}
\end{equation}
here we have coarsely sampled the training data every $0.02$ time units in order to reduce the amount of computation required by the regression.  The second term in the error function modifies ordinary linear least-squares regression in order to discourage overfitting; this modification is often called ridge regression or Tikhonov regularization.  Below, we will show results with regularization parameter $\beta = 10^{-6}$ and with $\beta = 0$ (no regularization).  We begin prediction by initializing $\hrr(T) = \rr(T)$ and evolving the predicting reservoir (\ref{eqn:predicting_reservoir}), where $T = 60$ is the end of the listening and training periods.

\begin{figure}[htbp]
	\centering
	\subfigure{\includegraphics[scale=.25]{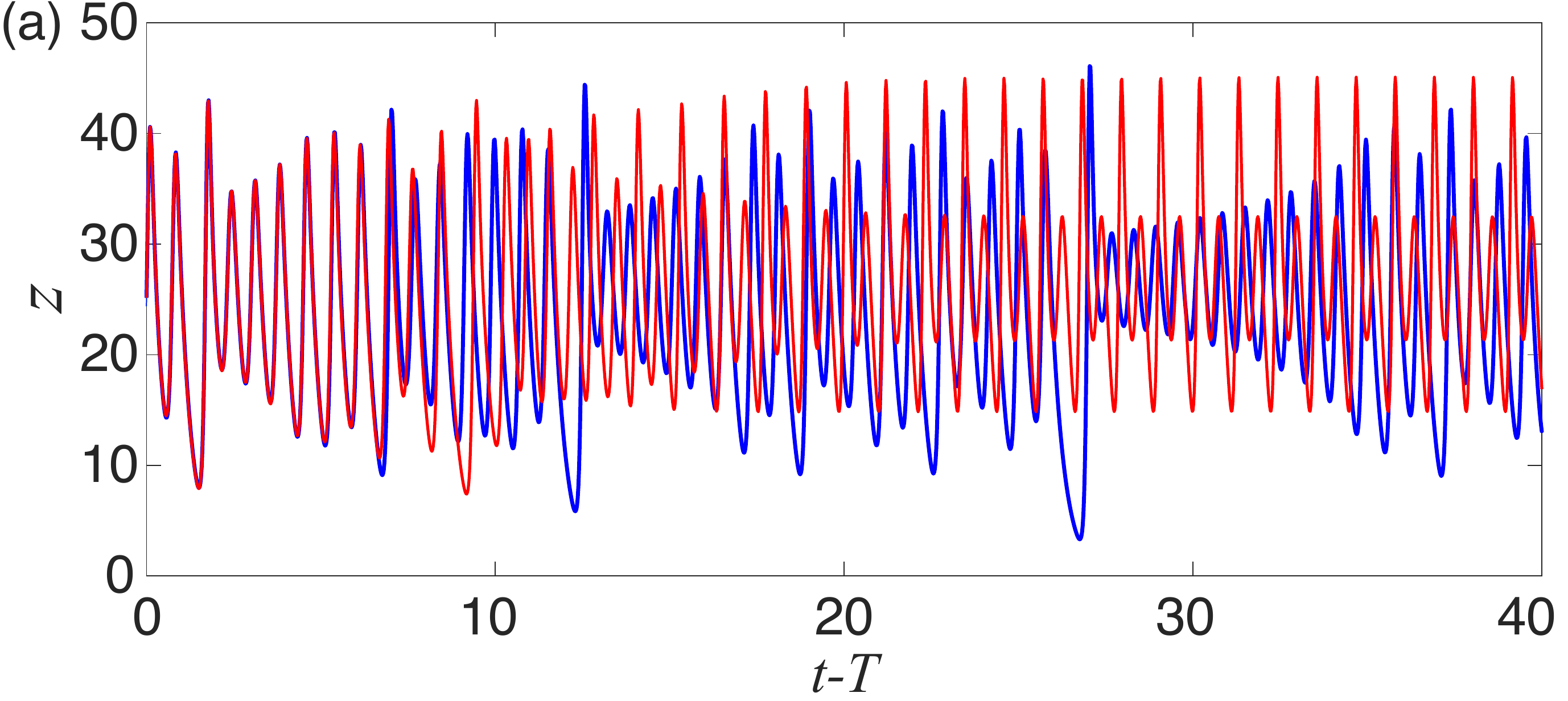}}
	\subfigure{\includegraphics[scale=.25]{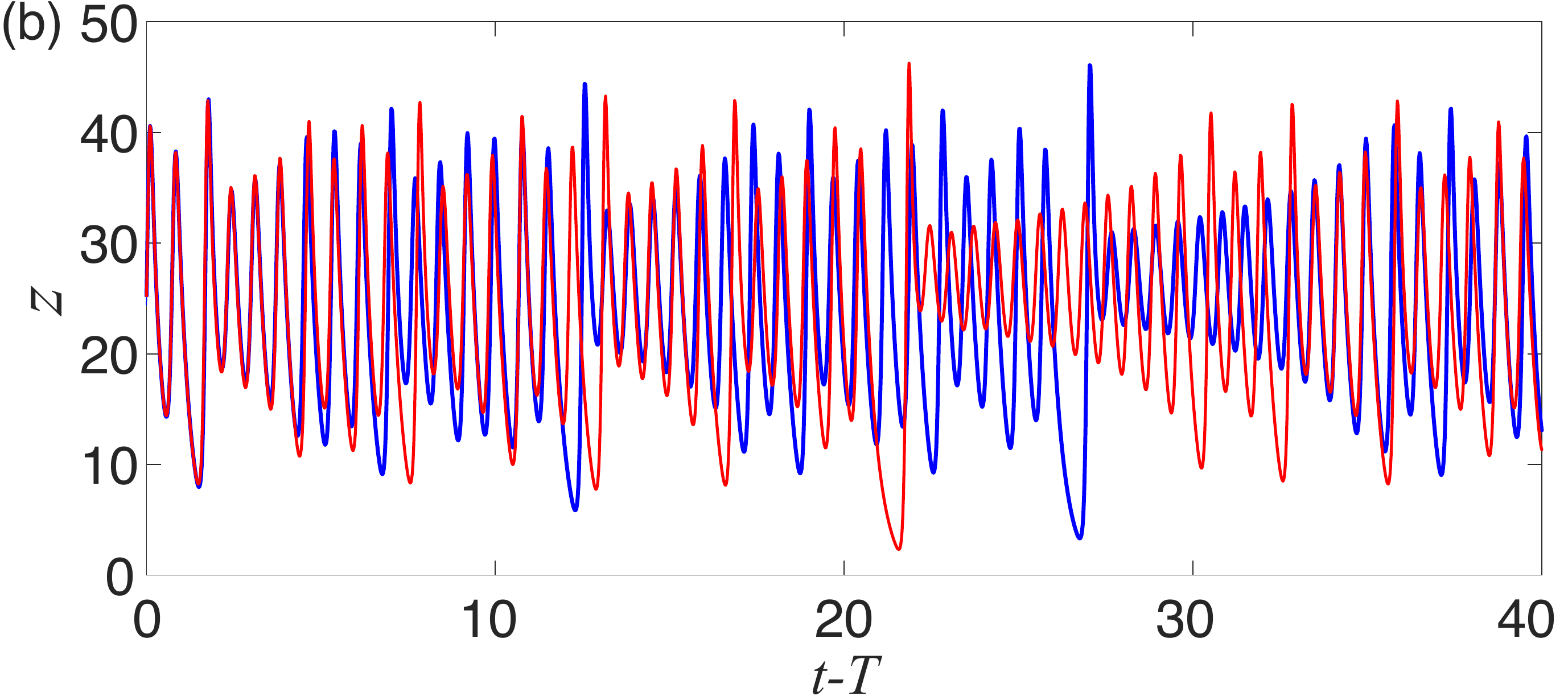}}
	\caption{Predicted (red) and actual (blue) $z(t)$ for a chaotic Lorenz system trajectory, using the same randomly-generated reservoir with different input strengths $\sigma=0.012$ [panel (a)] and $\sigma=0.014$ [panel (b)].  Both predictions remain well correlated with the actual trajectory for roughly $10$ time units.  After decorrelation, the first prediction approaches a periodic orbit, whereas the second prediction appears to continue with a climate similar to that of the actual trajectory.}
\label{fig:good_bad_prediction}
\end{figure}

\begin{figure}[htbp]
	\centering
	\subfigure{\includegraphics[scale=.22]{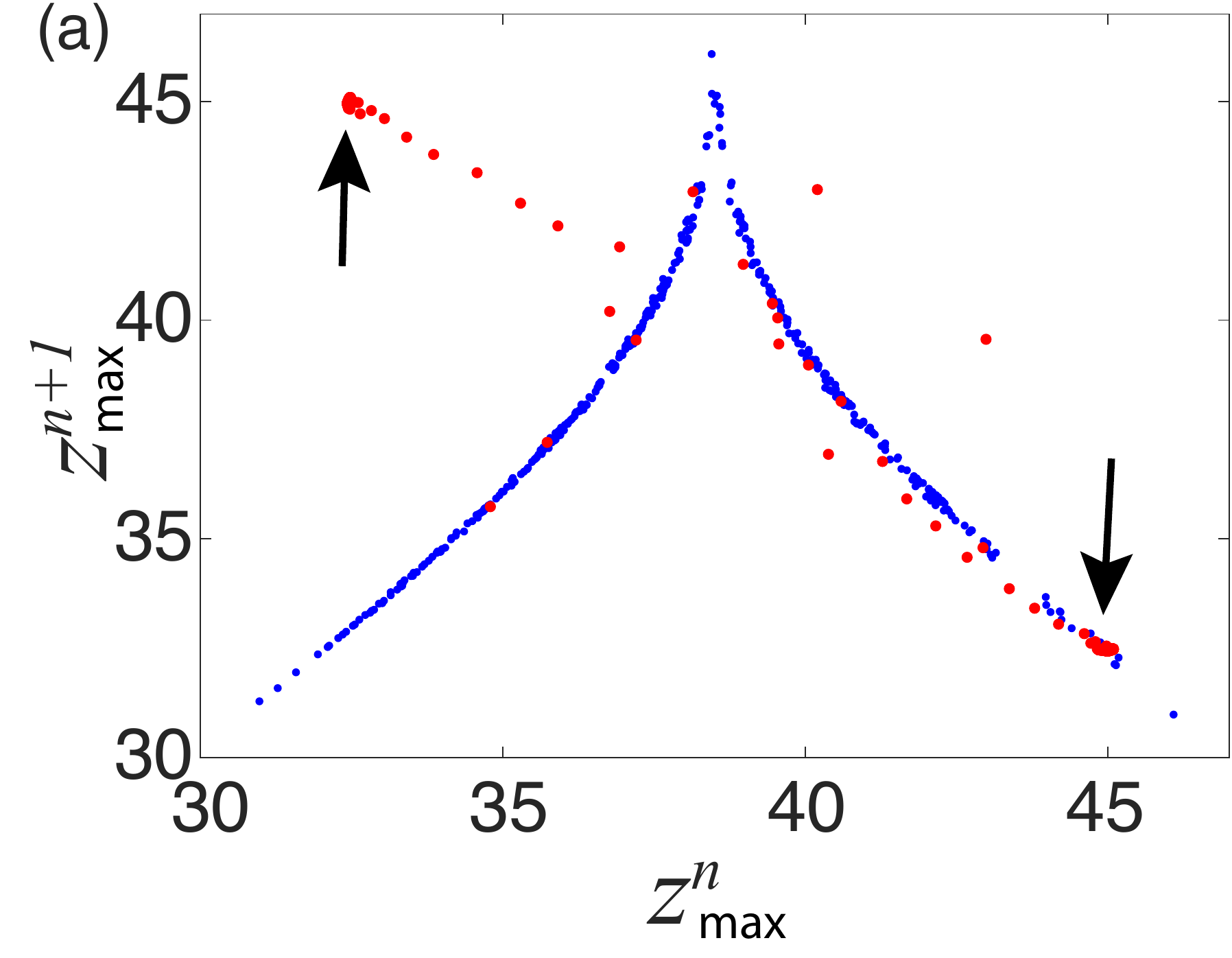}}
	\subfigure{\includegraphics[scale=.22]{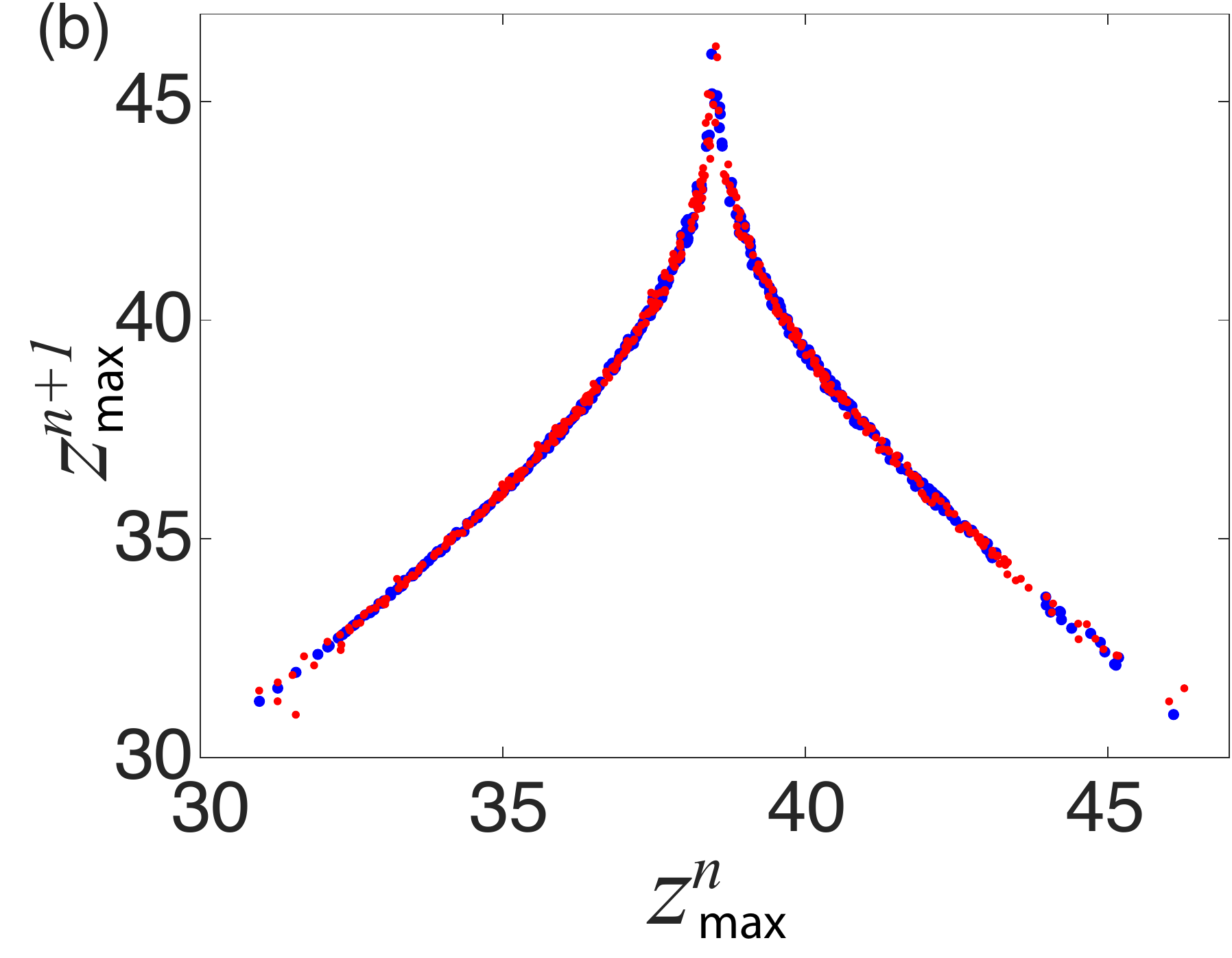}}
	\caption{Poincar\'e return map of successive local maxima of $z(t)$ for the actual (blue) and predicted (red) trajectories for $t-T$ from $0$ to $300$, using the same Lorenz trajectory and reservoir as Fig.~\ref{fig:good_bad_prediction}, again with $\sigma=0.012$ [panel (a)] and $\sigma=0.014$ [panel (b)].  Here $z^n_{\text{max}}$ represents the $n$th local maximum of $z(t)$.  The first prediction approaches a period two orbit (indicated by the arrows) that is not on the Lorenz attractor whereas the second prediction remains close to the Lorenz attractor.}
\label{fig:return_map}
\end{figure}

In Fig.~\ref{fig:good_bad_prediction}, we show the actual $z(t)$ from a trajectory of the Lorenz system, and predictions $\hat{z}(t)$ from two reservoirs that are identical except for their input strength parameter values [$\sigma=0.012$ for Fig.~\ref{fig:good_bad_prediction}(a) and $\sigma=0.014$ for Fig.~\ref{fig:good_bad_prediction}(b)].  Each reservoir is trained with the same Lorenz trajectory and with regularization parameter $\beta=10^{-6}$.  Both reservoirs predict the short-term future similarly well, but for larger values of the prediction time $t-T$, only the second prediction continues with a Lorenz-like climate.  We compare the two climate predictions over a longer time period in Fig.~\ref{fig:return_map}, which shows Poincar\'e return maps of successive $z(t)$ maxima.  In Fig.~\ref{fig:return_map}(a), the red dots (showing the reservoir prediction) initially are near the blue dots (representing the Lorenz attractor), but eventually the red dots approach a period two orbit, indicated by the arrows.  The large distance of the upper left arrow from the blue dots indicates that this period two orbit for the reservoir is not on the Lorenz attractor.  In contrast, the red dots in Fig.~\ref{fig:return_map}(b) remain near the blue dots at all times, indicating that the reservoir replicates the climate in the long term.

Based on the arguments in Sec.~\ref{sec:synch}, we hypothesize that for both $\sigma=0.012$ and $\sigma=0.014$, the listening reservoir (\ref{eqn:listening_reservoir}) evolves toward a set $\ph_\sigma(A)$, where $A$ is the Lorenz attractor and $\ph_\sigma$ is a generalized synchronization function.  Our choice of spectral radius $0.9$ for the adjacency matrix $\MM$ is consistent with common practice in reservoir computing \cite{caluwaerts2014}, though it does not guarantee uniform contraction for the listening reservoir \cite{yildiz2012}.  However, it does guarantee that the eigenvalues of the Jacobian matrix of the right side of (\ref{eqn:listening_reservoir}), evaluated at $\rr = \uu = 0$, have real parts at most $\gamma(-1+0.9) = 10(-0.1) = -1$.  This suggests an asymptotic contraction rate of $-1$ or faster for the listening reservoir, and that after discarding $100$ transient time units, $\rr(t)$ is extremely close to $\ph_\sigma(A)$ for $t \geq 0$.

Based on the arguments in Sec.~\ref{sec:attractor}, we hypothesize that the set $\ph_\sigma(A)$ is approximately invariant for the predicting reservoir (\ref{eqn:predicting_reservoir}).  Based on the results in Figs.~\ref{fig:good_bad_prediction} and \ref{fig:return_map}, we hypothesize further that for $\sigma=0.014$, there is an attracting invariant set for the predicting reservoir near $\ph_\sigma(A)$, but that between $\sigma=0.014$ and $\sigma=0.012$, there is a bifurcation that causes this invariant set either to become unstable or to be destroyed entirely.  To corroborate this hypothesis, we compute the Lyapunov exponents of the predicting reservoir for an approximate trajectory on $\ph_\sigma(A)$, as described in Sec.~\ref{sec:compute}.

\begin{figure}[htbp]
	\centering
	\includegraphics[scale=.3]{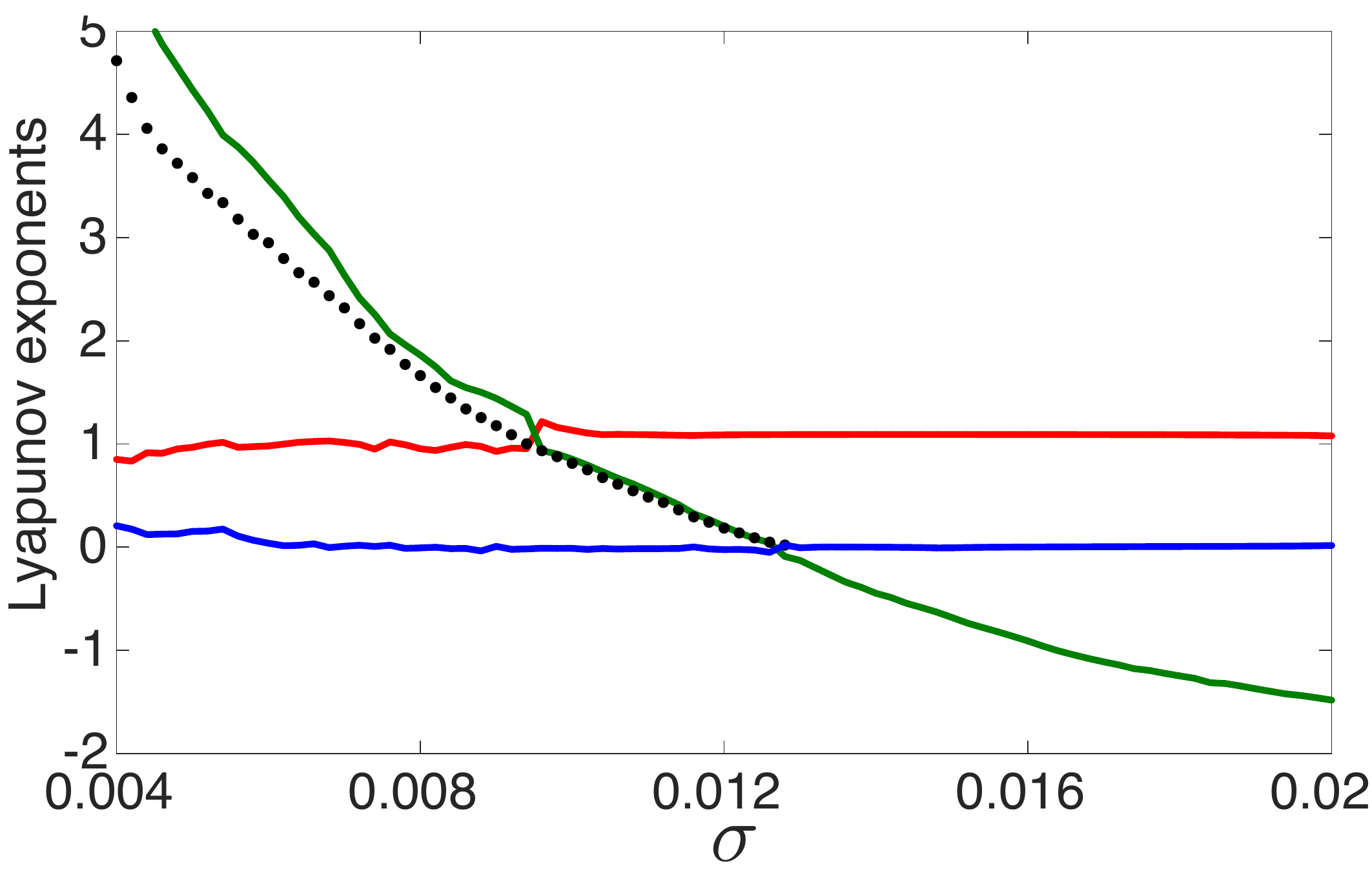}
	\caption{The three largest Lyapunov exponents of the predicting reservoir (\ref{eqn:predicting_reservoir}) on the invariant set $\ph_\sigma(A)$ for the listening reservoir (\ref{eqn:listening_reservoir}), as a function of the input strength $\sigma$, for the same reservoir as Figs.~\ref{fig:good_bad_prediction} and \ref{fig:return_map}.  Two exponents that are approximately constant as a function of $\sigma$, and which approximate the two largest Lyapunov exponents of the Lorenz attractor, are colored red and blue; the more variable exponent, which we call the transverse Lyapunov exponent and which determines climate stability, is colored green.  For values of $\sigma$ for which we detect divergence from the Lorenz climate, we graph with a black dot the observed divergence rate $\lambda^*$, computed as described in the text.}
\label{fig:Lyp_ridge_regression}
\end{figure}

Fig.~\ref{fig:Lyp_ridge_regression} shows the three largest Lyapunov exponents of the predicting reservoir (\ref{eqn:predicting_reservoir}) as the input strength $\sigma$ varies from $0.004$ to $0.02$.  We do not change the matrices $\MM$ and $\Win$, but for each value of $\sigma$, we perform a separate training (with $\beta = 10^{-6}$ as before), resulting in a different output weight matrix $\Wout$.  The exponents colored red and blue approximate the positive and zero Lyapunov exponents of the Lorenz attractor $A$ (the approximation is closest for $\sigma \geq 0.01$).  Reproduction of the positive exponent of $A$ in the reservoir dynamics on $\ph_\sigma(A)$ is a necessary consequence of successful attractor reconstruction, and does not indicate instability of $\ph_\sigma(A)$ to transverse perturbations.  The exponent colored green estimates the largest of the transverse Lyapunov exponents described in Sec.~\ref{sec:lyapunov}.  This exponent passes through zero, indicating a bifurcation, at $\sigma\approx 0.013$.

Next, we compare the change in stability indicated by the computed transient Lyapunov exponent to a more direct computation indicating success or failure of climate replication.  To detect when the prediction $\huu(t) = \hps(\hrr(t))$ of the Lorenz state diverges from the true Lorenz attractor, we let $\Delta(t)$ be the Euclidean distance between the vector field $d\huu/dt$ implied by the predicting reservoir and the vector field (right-hand side) of the Lorenz system (\ref{eqn:lorenz}), evaluated at $[x,y,z]^T = \huu(t)$. [We calculate the reservoir-implied vector field by the chain rule $d\huu/dt = D\hps(\hrr(t)) d\hrr/dt$, where $D\hps$ is the Jacobian matrix of $\hps = \Wout\qq$, and $d\hrr/dt$ is given by Eq.~(\ref{eqn:predicting_reservoir}).]  For each value of $\sigma$ depicted in Fig.~\ref{fig:Lyp_ridge_regression}, we calculate the vector field discrepancy $\Delta(t)$ for the prediction time period $t \geq T$. If $\Delta(t)$ does not exceed a threshold value $20$ for a duration of $800$ time units, we consider the climate to be approximately reproduced. (Our threshold value $20$ is small compared to the typical magnitude of the Lorenz vector field.)  Otherwise, we say that the prediction has ``escaped'' from the Lorenz attractor.  In Fig.~\ref{fig:Lyp_ridge_regression}, we show a black dot at each value of $\sigma$ for which we detect escape; these values are the same as those for which the computed transverse Lyapunov exponent is positive.  The height of each black dot represents an observed divergence rate $\lambda^*$, computed as follows.

When we detect escape for a particular value of $\sigma$, we reinitialize the predicting reservoir (\ref{eqn:predicting_reservoir}) using $\hrr(t_0) = \rr(t_0)$ for $1000$ different values of $t_0 \geq T$, where the values of $\rr(t_0)$ are determined by continuing to run the listening reservoir (\ref{eqn:listening_reservoir}) for $t \geq T$.  For each $t_0$, we evolve the predicting reservoir until the first time $t_1$ for which $\Delta(t_1) \geq 20$, or until $t_1-t_0 = 800$, whichever comes first.  If divergence from the attractor is governed by Lyapunov exponent $\lambda$, we should have
$\Delta(t_1) \approx \Delta(t_0) \exp(\lambda (t_1-t_0))$ in a certain average sense.  We compute the observed exponential divergence rate $\lambda^*=\langle \ln[\Delta(t_1)/\Delta(t_0)] \rangle / \langle t_1-t_0\rangle$, where the angle brackets represent an average over the $1000$ values of $t_0$.
The computed values of $\lambda^*$ are shown as black dots in Fig.~\ref{fig:Lyp_ridge_regression}. The approximate agreement of $\lambda^*$ with the green curve (especially for $0.01 \leq \sigma \leq 0.013$) demonstrates that the computed transverse Lyapunov exponent reflects divergence of predictions from the Lorenz attractor. 

\begin{figure}[htbp]
	\includegraphics[scale=.35]{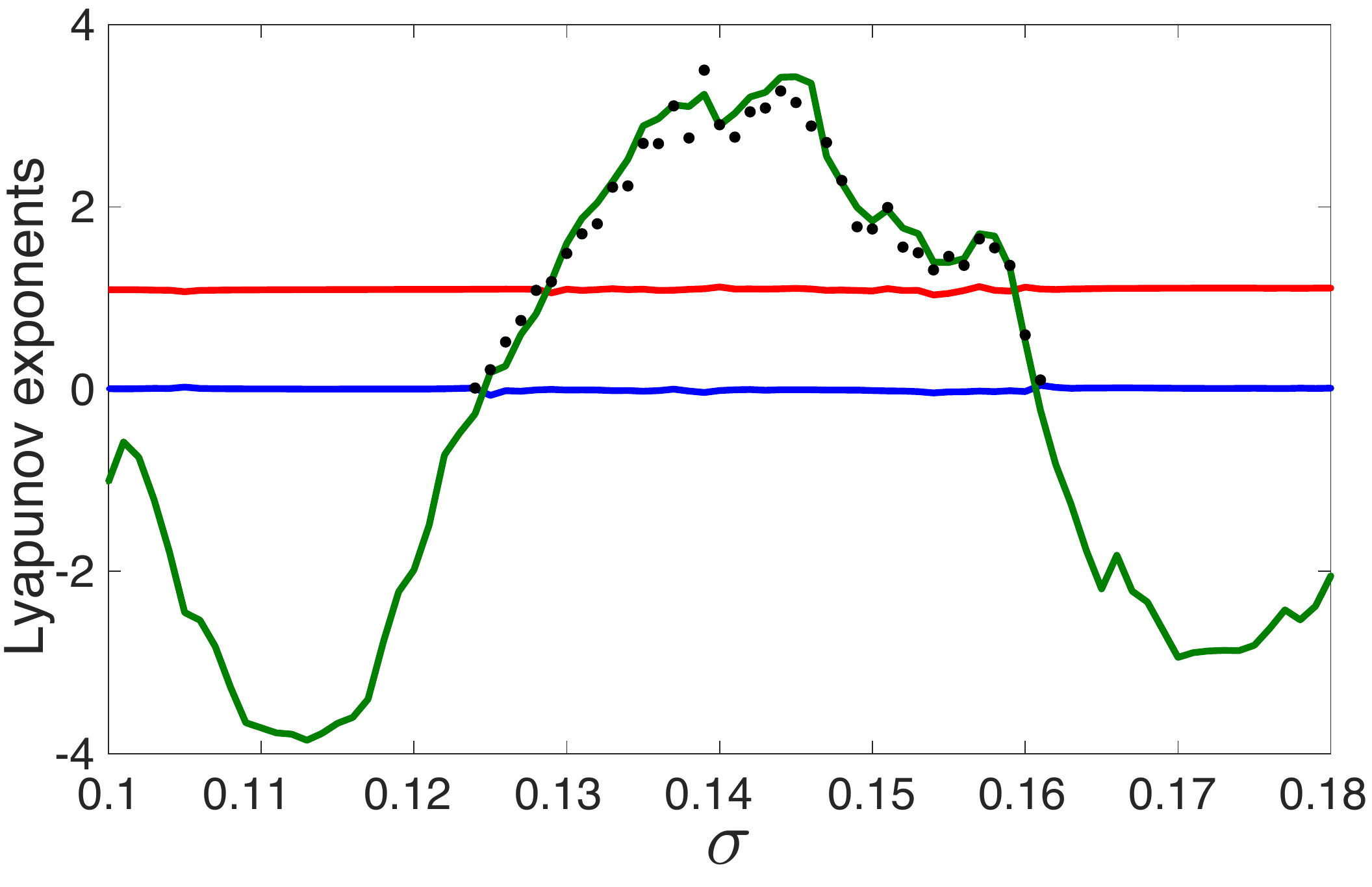}
	\caption{The three largest Lyapunov exponents of the predicting reservoir (\ref{eqn:predicting_reservoir}), and the estimated divergence rate $\lambda^*$, as a function of $\sigma$, using the same color scheme as Fig.~\ref{fig:Lyp_ridge_regression}.  Here we use a different randomly-generated reservoir than in Fig.~\ref{fig:Lyp_ridge_regression}, and no regularization ($\beta = 0$) in the training.}
\label{fig:LYP}
\end{figure}

To illustrate the correspondence between the computed transverse Lyapunov exponent and the observed divergence rates in a case where their dependence on $\sigma$ is more complicated, we show in Fig.~\ref{fig:LYP} the analogue of Fig.~\ref{fig:Lyp_ridge_regression} in a case where no regularization  ($\beta = 0$) is used in the training.  Again, we see precise correspondence between detected failure of climate replication (presence of a black dot) and positive values of the transverse Lyapunov exponent (green curve), and good agreement with the observed divergence rates for these values of $\sigma$.  In this case, there are two bifurcations, one near $\sigma = 0.12$ and one near $\sigma = 0.16$.

\begin{figure}[htbp]
	\centering
	\subfigure{\includegraphics[scale=.28]{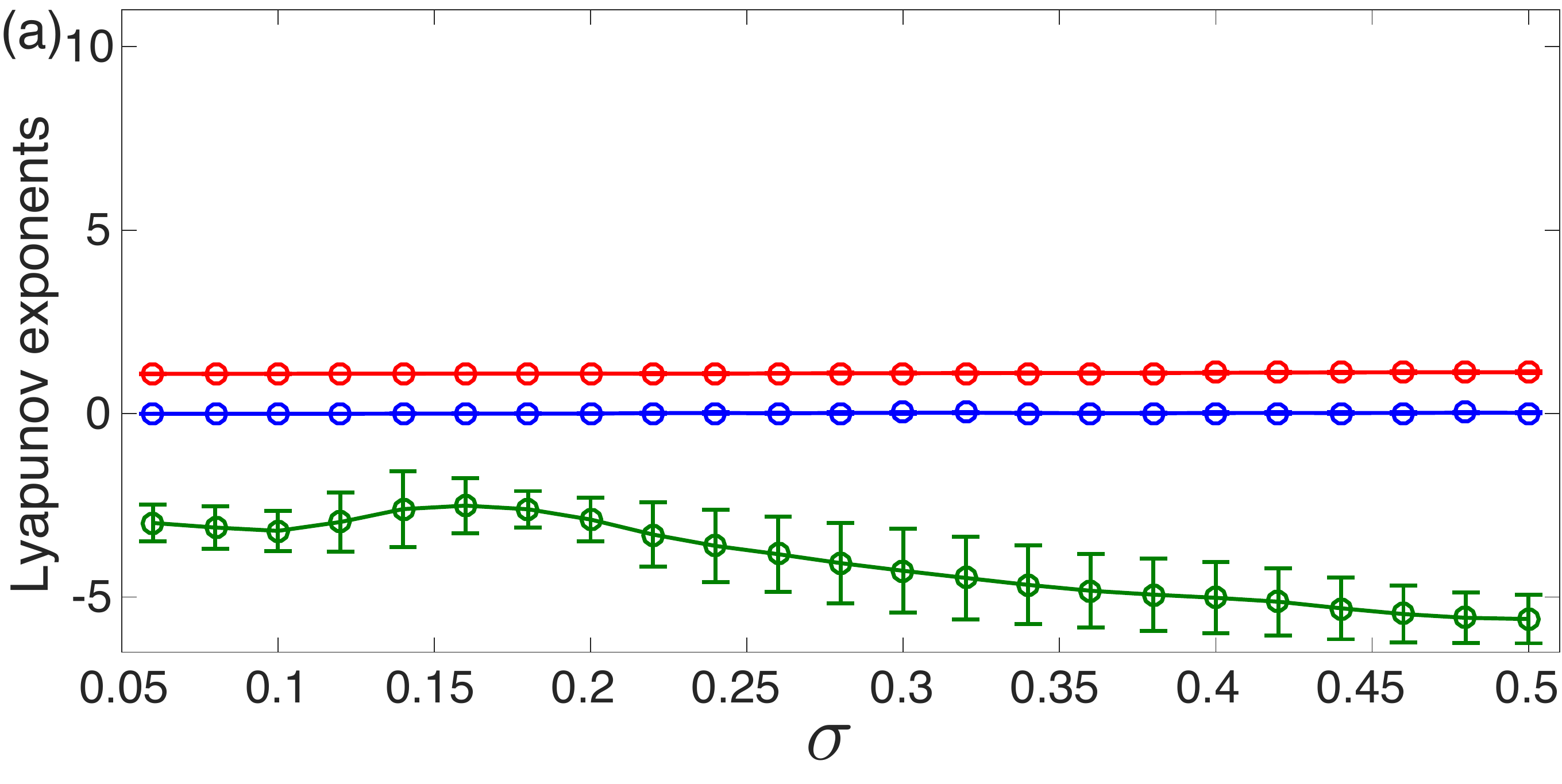}}\\
	\subfigure{\includegraphics[scale=.28]{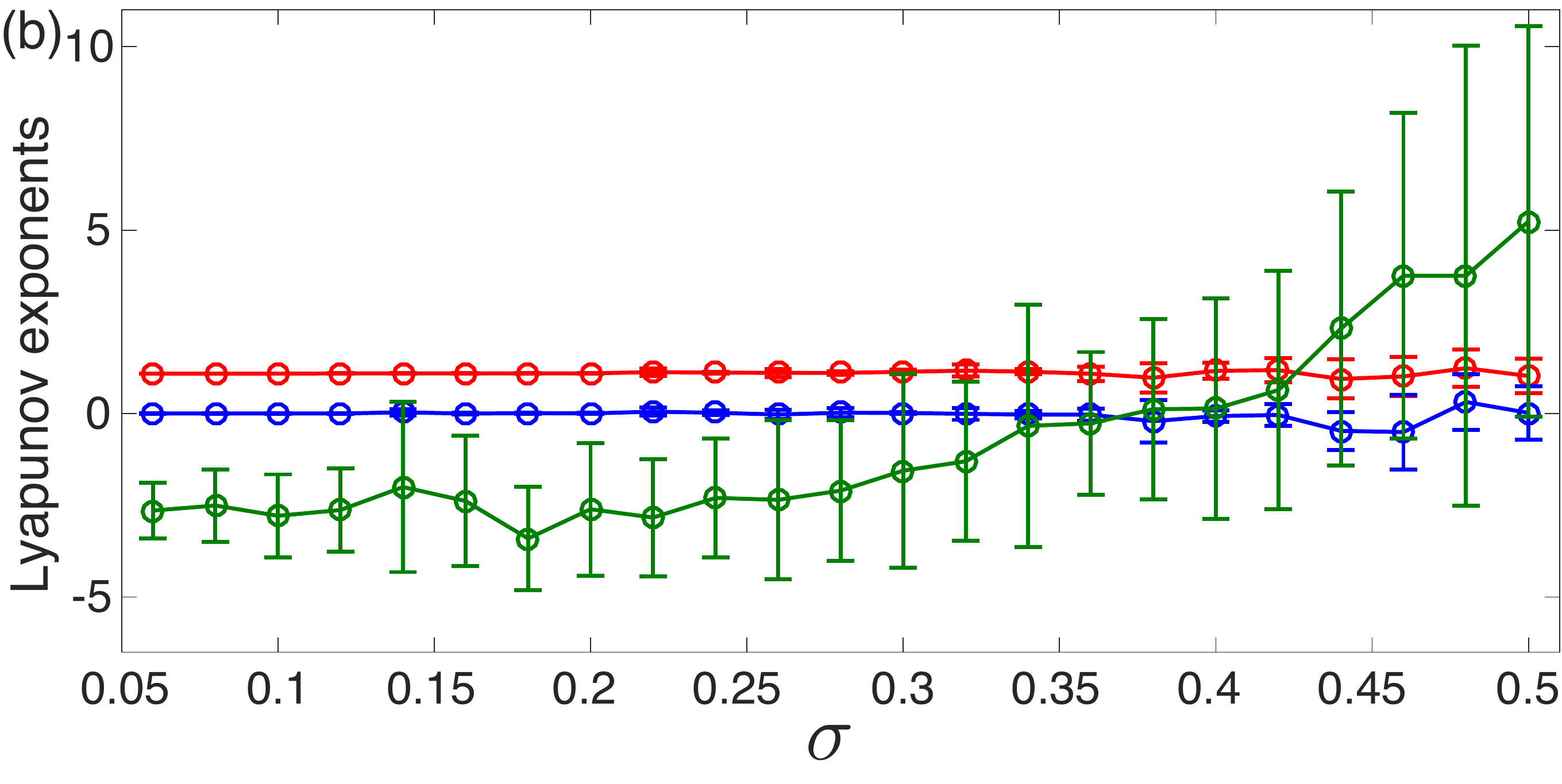}}
	\caption{The means and standard deviations of three largest Lyapunov exponents for the same $10$ randomly-generated reservoirs trained with regularization parameter $\beta = 10^{-6}$ [panel (a)] and with $\beta=0$ [panel (b)].  Again, the red and blue curves approximate the two largest exponents of the Lorenz attractor, and the green curve is the computed transverse Lyapunov exponent.}
\label{fig:LYP_ridge_and_no_ridge}
\end{figure}

We remark that when we use regularization ($\beta = 10^{-6}$) in the training, we do not observe as complicated a dependence of the computed transverse Lyapunov exponent on the input strength $\sigma$ as in Fig.~\ref{fig:LYP}.  Instead, the computed transverse Lyapunov exponent is typically negative and slowly varying across a wide range of $\sigma$ values, for which climate replication is successful.  In Fig.~\ref{fig:LYP_ridge_and_no_ridge}, we use the transverse Lyapunov exponent computation, averaged over $10$ different randomly-generated reservoirs, to give a quantitative illustration of the advantage of regularization.  When regularization is used, the negative means and small standard deviations of the computed transverse Lyapunov exponent indicate robust climate stability over the entire range $0.05 \leq \sigma \leq 0.5$.  (By contrast, Figs.~\ref{fig:good_bad_prediction}--\ref{fig:Lyp_ridge_regression} depicted values of $\sigma \leq 0.02$.)  With no regularization, the means are larger and more variable, indicating less stability and greater sensitivity to the value of $\sigma$, and the standard deviations are significantly larger, indicating lack of robustness from one random reservoir realization to another.

\section{Conclusions and Discussion}
\label{sec:discussion}

We presented in Sec.~\ref{sec:Hunt} a partial explanation for how reservoir computing prediction is able to reconstruct the attractor (replicate the climate) for a chaotic process from limited time series data.  We argued that the reservoir dynamics  (\ref{eqnon}) can be designed so that during the listening period on which training is based, the reservoir state $\rr(t)$ is approximately a continuous function $\ph$ of the state $\sss(t)$ of the chaotic process.  This property, called generalized synchronization, is closely related to the echo state property for reservoir computing.  We showed that both properties hold if the listening reservoir (\ref{eqnon}) is uniformly contracting as a function of the reservoir state; other criteria for these properties have also been identified \cite{stark1997,yildiz2012,caluwaerts2014}.

Ideally, the synchronization function $\ph$ should be one-to-one in order to recover the process dynamics from the reservoir dynamics.  Investigation of conditions that can guarantee $\ph$ to be one-to-one could help guide reservoir design.  However, even in the absence of a guarantee, we noted that embedding results suggest that $\ph$ is likely to be one-to-one if the reservoir state space is sufficiently high-dimensional compared with dimensionality of the chaotic process.

Practically speaking, a necessary condition for climate replication is that training be successful in approximately recovering the measured state $\uu(t) = \hh(\sss(t))$ from the reservoir state $\rr(t)$; this depends on the amount of training data available and the method of regression used, among other things.  We did not address theoretical aspects of training, but we argued that success is plausible if the reservoir is sufficiently high-dimensional and heterogeneous to yield a large variety of basis functions for the regression.

We showed that in the limit that the approximations we described are exact, the predicting reservoir (\ref{eqaut}) exactly predicts future values of $\uu(t)$.  Thus, accurate approximations yield commensurately accurate short-term forecasts.  Long-term climate replication depends on stability of the predicting reservoir dynamics with respect to perturbations produced by the approximations.  We discussed how to estimate Lyapunov exponents for the predicting reservoir in numerical experiments, whether or not the desired climate is stable.  We emphasize that our computation of Lyapunov exponents was intended to illustrate our theory, and that the method we described requires measurements $\{\uu(t)\}$ over a long time period to maintain the desired climate.  If one's goal is to estimate the Lyapunov exponents of the process that produced $\{\uu(t)\}$ from a limited amount of data, one should seek parameters of the predicting reservoir that replicate the climate, and simply compute the Lyapunov exponents of the resulting trajectory\cite{pathak2017using}.

In Sec.~\ref{sec:RCN}, we gave examples of climate replication successes and failures, and showed how they correspond to the Lyapunov exponents we computed.  We emphasize that the results and the ranges of $\sigma$ we displayed were selected to illustrate and analyze failures that can occur with inadequate input strength (Figs.~\ref{fig:good_bad_prediction}--\ref{fig:Lyp_ridge_regression}) or without regularization (Fig.~\ref{fig:LYP}) in the training.  With regularization, we are able to obtain robust climate replication [indicated by Fig.~\ref{fig:LYP_ridge_and_no_ridge}(a)] over a wide range of input strengths.

We remark that for simplicity, our theory considered discrete-time reservoir dynamics.  Discrete time is the appropriate way to model software reservoirs, but physical reservoirs typically are better modeled by continuous time.  With appropriate modifications, our theory applies to the continuous-time case.   The prediction time increment $\tau$ used in the training should be the amount of time information takes to traverse the feedback loop depicted in Fig.~\ref{fig:reservoir2}.  However, with a physical reservoir, careful calibration of the sampled training data may be necessary to meet the goal of predicting $\uu(t+\tau)$ based on the listening reservoir's response to input up to time $t$, in part because $\tau$ is a property of the predicting reservoir and not of the listening reservoir.

Finally, we argue that in addition to reservoir computing, the theory we presented in Section~\ref{sec:Hunt} applies to some other machine learning methods for time series prediction.  The essential features a prediction method needs for our theory to apply are:  (1) that the method maintains an internal state, or ``memory'', that depends on the sequence of inputs it receives during training; (2) that it is trained to predict a short time increment ahead, after receiving the input time series for a relatively long time interval; and (3) that it is used to predict farther into the future by iterating its incremental forecasts through a feedback loop.  These features are present, for example, in prediction using the FORCE method for training reservoirs \cite{sussillo2009generating} and in recent work using long short-term memory (LSTM) networks for prediction \cite{vlachas2018}.  For methods that (unlike reservoir computing) train parameters that affect the internal state in the absence of feedback, our theory applies if we take the function $\fff$ in Eq.~(\ref{eqnon}) to represent the update rule for the internal state $\rr$ after training has selected parameter values.  Though our description of how training arrives at the pair of functions ($\fff$,$\hps$) was specific to reservoir computing, our discussion of how these functions can be used with Eqs.~(\ref{eqaut}) and (\ref{eqpred}) for prediction and attractor reconstruction are independent of which machine-learning method is used to determine the functions.

We gratefully acknowledge the support of grants from ARO (W911NF-12-1-0101) and DARPA. We thank Michelle Girvan, Jaideep Pathak, Sarthak Chandra, Daniel Gauthier, and Daniel Canaday for their input.
\bibliographystyle{aipnum4-1}
\bibliography{citation}

\end{document}